\documentclass[a4paper,12pt]{article}
\usepackage[utf8]{inputenc}
\usepackage{ucs}
\usepackage[english]{babel}
\usepackage[numbers,comma,square,sort,compress]{natbib}

\usepackage[a4paper,hmargin=26mm,vmargin=24mm]{geometry}
\usepackage{slashed,braket}
\usepackage{subcaption}
\usepackage{amsmath,amssymb,amsfonts}
\usepackage{hyperref}

\usepackage{tikz-feynman}
\tikzfeynmanset{compat=1.1.0}

\DeclareMathOperator{\Real}{Re}
\DeclareMathOperator{\Imag}{Im}

\newcommand{\emailfn}[1]{\footnote{\href{mailto:#1}{#1}}}

\begin{document}
\begin{center}
{\Large{\bf Renormalizable and unitary nonlocal quantum field\\\vspace{.3em} theory with CPT violation and its implication}${}^\dagger$}
\end{center}
\vspace{.5em}
\begin{center}
{\bf {Moshe M. Chaichian,\emailfn{masud.chaichian@helsinki.fi} Markku A. Oksanen,\emailfn{markku.oksanen@helsinki.fi} and Anca Tureanu\,\emailfn{anca.tureanu@helsinki.fi}}}
\end{center}
\vspace{.2em}
\begin{center}
{\it Department of Physics, University of Helsinki,\\ 
P.O.Box 64, FI-00014 University of Helsinki, Finland\\
\vspace{.2em}
and\\
\vspace{.3em}
Helsinki Institute of Physics,\\ 
P.O.Box 64, FI-00014 University of Helsinki, Finland}
\end{center}
\vspace{.2em}

\begin{abstract}
It is a common belief that any relativistic nonlocal quantum field theory encounters either the problem of renormalizability or unitarity or both of them. It is also known that any local relativistic quantum field theory (QFT) possesses the CPT symmetry. In this Letter we show that a previously proposed nonlocal Lorentz invariant QFT, which   violates the CPT theorem, is both renormalizable and unitary, thus being a first presented example in the literature of such a nonlocal theory. The theory satisfies the requirement of causality as well. A further generalization of such a nonlocal QFT to include the gauge theories is also envisaged. In particular, dressing such a Standard Model with a CP violating phase, will make the theory satisfying most of the necessary criteria to finally explain the baryon asymmetry of the universe by a viable QFT. As for the necessity of baryon number violation, there are hopefully several possibilities such as by GUT and electroweak baryogenesis, leptogenesis or sphalerons.
\end{abstract}
\vspace{.2em}


\begin{center}
\noindent${}^\dagger$\textit{Dedicated to the warm memory of Matts Roos (28 October 1931 -- 25 November 2025, Helsinki), a dear Colleague and Friend.}
\end{center}

\section{Introduction}
We consider nonlocal quantum field theories which are Lorentz invariant but violate CPT invariance \cite{Luders:1954zz,Pauli:1955}. Such theories have been used to study the consequences of CPT violation without assuming that Lorentz invariance is violated \cite{Chaichian:2011fc,Duetsch:2012sd,Chaichian:2012ga}.
It has been shown that CPT violation in Lorentz invariant quantum field theory has distinct effects, depending on whether the CPT violation is implemented in a mass term or in an interaction term \cite{Alvarez-Gaume:2023few}.
The spin-statistics theorem \cite{Pauli:1940zz,Luders:1958zz,Burgoyne:1958} was also shown to remain valid in theories of this type \cite{Alvarez-Gaume:2023few}.
In this letter, we show that theories of this type are both renormalizable and perturbatively unitary.

There exists no other example of a viable QFT that breaks CPT but is unitary, regardless of whether Lorentz invariance is violated or not.

Theories nonlocal in time are known to lack an unambiguous definition of canonical momenta and consequently canonical quantization. 
A theory describing a nonlocal Yukawa coupling was proposed in \cite{Kristensen:1952}. Nonlocal field theories were studied in detail by Marnelius \cite{Marnelius:1973iw,Marnelius:1974rq}, using a procedure close to canonical quantization, who showed that models like that of \cite{Kristensen:1952} are inconsistent, since energy-momentum conservation is broken down and a unitary $S$ operator cannot be defined. Our approach differs from the above mentioned in that nonlocal contributions to interactions are perturbations to local interactions (and similarly for mass terms). Furthermore, quantization is based on Schwinger’s action principle \cite{Schwinger:1951nm,Schwinger:1951xk} (see also \cite{Fujikawa:2004rt,Chaichian:2012ga}).

Renormalization of local quantum field theories is well understood \cite{Matthews:1951sk,Matthews:1954sf,Bogoliubov:1955,Bogoliubov:1957,Hepp:1966eg,Zimmermann:1969jj}. For more detailed presentations as book material, we refer to \cite{Bogoliubov:1959,Weinberg:1995mt,Nishijima:2023}. The situation with nonlocal quantum field theories is more complex, since there are several ways to construct nonlocal theories. 
We emphasize that by nonlocal theory we mean that the action is nonlocal but not due to higher derivatives of the fields.
For some recent attempts to renormalize different nonlocal theories, see \cite{Thurigen:2021zwf,Abu-Ajamieh:2023syy}.

We stress that renormalizability and perturbative unitarity are independent properties, in the sense that one does not imply the other \cite{Scharf:1995,Duetsch:2019}.

\section{The model}
For simplicity we consider a theory of a single spin-1/2 field $\psi$ and a single scalar field $\phi$ with Yukawa coupling.
We implement the Lorentz invariant CPT violation with nonlocal terms which feature the kernel
\begin{equation}\label{Fxy}
F(x,y)=\theta(x^0-y^0)\delta((x-y)^2-l^2),
\end{equation}
where $\theta$ is the Heaviside step function for the difference of temporal coordinates $x^0-y^0$ for two points in spacetime and the real parameter $l$ is the scale of the nonlocal interaction \cite{Chaichian:2012ga}. 
The kernel is chosen to fulfill the requirements of causality and invariance under translations and Lorentz transformations. The combination \eqref{Fxy}, with the Heaviside step function having values $0$ and $1$, ensures Lorentz invariance, i.e. invariance under the proper orthochronous Lorentz transformations \cite{Chaichian:2011fc}. Causality is ensured by \eqref{Fxy}, since a nonlocal interaction term has support only inside the light cone, so that the (anti)commutation relations vanish outside of the light cone \cite{Bogoliubov:1959,Weinberg:1995mt,Nishijima:2023}. 
Description of the causality of measurements in a relativistic quantum field theory, particularly for observables extended not only in space but also in time, is a challenge that has seen recent advances \cite{Oeckl:2025thz}. 
Concerning the range of nonlocality, instead of the fixed spacetime interval by $\delta((x-y)^2-l^2)$ we could use in \eqref{Fxy}, for example, $\theta((x-y)^2)e^{-(x-y)^2/l^2}$ for all causally connected points \cite{Chaichian:2011fc}.

Consider the following Lagrangian
\begin{equation}\label{Lagrangian}
\mathcal{L}=\mathcal{L}_\psi+\mathcal{L}_\phi+\mathcal{L}_\mathrm{int},
\end{equation}
where the free action of the fermion $\psi$ includes a nonlocal mass term that produces particle-antiparticle mass splitting,
\begin{equation}\label{Lagrangian.psi}
\begin{split}
\mathcal{L}_\psi&=\bar{\psi}(x)i\gamma^{\mu}\partial_{\mu}\psi(x) - m\bar{\psi}(x)\psi(x)\\
&\quad -\mu\int d^{4}y \left[\theta(x^{0}-y^{0})-\theta(y^{0}-x^{0})\right] \delta((x-y)^2-l^2)\,i\bar{\psi}(x)\psi(y).
\end{split}
\end{equation}
The free action of the real scalar field $\phi$ is the usual local one
\begin{equation}\label{Lagrangian.phi}
\mathcal{L}_\phi=\frac{1}{2}\partial^\mu\phi(x)\partial_\mu\phi(x)-\frac{1}{2}m_{\phi}^2\phi^2(x),
\end{equation}
and the interaction terms contain a local Yukawa interaction, a nonlocal contribution to the Yukawa interaction, as well as a local $\phi^4$-self-interaction for the scalar field,
\begin{multline}\label{Lagrangian.int}
\mathcal{L}_\mathrm{int}= g\bar\psi(x)\psi(x)\phi(x) +g_1\bar\psi(x)\psi(x)
\int d^4y \theta(x^0-y^0)\delta((x-y)^2-l^2)\phi(y)\\ -\frac{\lambda}{4!}\phi^4(x).
\end{multline}
The theory is Lorentz invariant but violates CPT due to the nonlocal fermion mass term in \eqref{Lagrangian.psi} and the nonlocal part of the Yukawa interaction \eqref{Lagrangian.int}.
The scalar self-interaction term $\phi^4$ is included in order to have a renormalizable local theory; no nonlocal $\phi^4$-term is included here, since it would not introduce anything qualitatively new to the theory.
The corresponding local theory is known to be renormalizable and perturbatively unitary \cite{Matthews:1951sk,Matthews:1954sf}. In the following, we show that the nonlocal theory retains those properties.

The effects of the CPT-violating mass and interaction terms can be described in momentum space by the following two form factors \cite{Chaichian:2012ga}:
\begin{equation}\label{formfactors}
f_{\pm}(k)=\int d^4z\, \theta(z^0)\delta(z^2-l^2)e^{\pm ik\cdot z}.
\end{equation}
The propagator of the fermion in momentum space is
\begin{equation}\label{propagator.psi}
S_F(p)=\frac{i}{\slashed{p}-m+\mu \Delta f(p)+i\epsilon},
\end{equation}
where we denote the difference of the two form factors \eqref{formfactors} as
\begin{equation}
\begin{split}
\Delta f(k)&=-i\left[f_+(k)-f_-(k)\right]\\
&=2\int d^4z\,\sin(k\cdot z)\theta(z^0)\delta(z^2-l^2).
\end{split}
\end{equation}
The propagator of the scalar field is the same one as in a local theory
\begin{equation}
D_F(k)=\frac{i}{k^2-m_{\phi}^2+i\epsilon}.
\end{equation}

\begin{figure}[ht]
\centering
\begin{subfigure}{0.4\textwidth}
  \centering
  \begin{tikzpicture}
    \begin{feynman}[small]
      \vertex (a);
      \vertex [right=of a, dot, label={[right=2mm]{$-i\left[g+g_1f_+(k)\right]$}}] (b) {};
      \vertex [above right=of b] (c);
      \vertex [below right=of b] (d);
      \diagram* {
        (a) -- [fermion] (b)
        -- [fermion] (c),
        (d) -- [charged scalar, edge label=\(k\)] (b),
      };
    \end{feynman}
  \end{tikzpicture}
  \caption{}\label{fig:1a}
\end{subfigure}
\begin{subfigure}{0.4\textwidth}
  \begin{tikzpicture}
    \begin{feynman}[small]
      \vertex (a);
      \vertex [right=of a, dot, label={[right=2mm]{$-i\left[g+g_1f_-(k)\right]$}}] (b) {};
      \vertex [above right=of b] (c);
      \vertex [below right=of b] (d);
      \diagram* {
        (a) -- [fermion] (b)
        -- [fermion] (c),
        (b) -- [charged scalar, edge label'=\(k\)] (d),
      };
    \end{feynman}
  \end{tikzpicture}
  \caption{}\label{fig:1b}
\end{subfigure}
\caption{Feynman rules for the fermion-scalar vertex of the nonlocal Yukawa interaction \eqref{Lagrangian.int}: (a) absorption of a scalar particle, (b) creation of a scalar particle.}
\label{fig:treediagrams}
\end{figure}
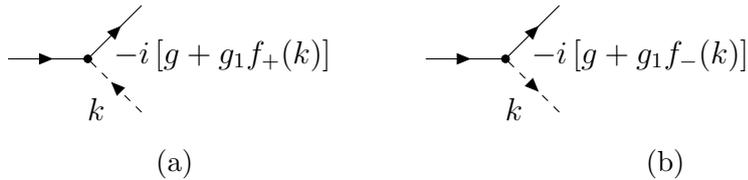

The action of the nonlocal part of the Yukawa interaction can be written in momentum space as
\begin{multline}
g_1\int d^4x\bar\psi(x)\psi(x)\int d^4y \theta(x^0-y^0)\delta((x-y)^2-l^2)\phi(y)\\
=g_1\int d^4p_1d^4p_2d^4k(2\pi)^4\delta^4(p_1+p_2+k)
\bar\psi(p_1)\psi(p_2)f_+(k)\phi(k).
\end{multline}
For the $\bar\psi\psi\phi$-vertex, where a scalar particle of momentum $k$ is annihilated, the Feynman rule is $-i\left[g+g_1f_+(k)\right]$, while for a vertex, where a scalar particle of momentum $k$ is created, the Feynman rule is $-i\left[g+g_1f_-(k)\right]$ (see Fig.~\ref{fig:treediagrams}).

\section{Renormalizability}
We remark on the dimensions of the coupling constants of the nonlocal terms in the Lagrangian \eqref{Lagrangian}--\eqref{Lagrangian.int}. We use natural units, $\hbar=c=1$, so that length and time have the dimension of mass to power $-1$. The mass-splitting coupling $\mu$ has mass dimension $3$, and the nonlocal Yukawa coupling $g_1$ has mass dimension $2$. Generally, since an integral involving \eqref{Fxy} adds a mass dimension $-2$, the coupling constant of a nonlocal term has a mass dimension two higher than the coupling of the corresponding local term. Thus, for any renormalizable local interaction, the nonlocal couplings have positive mass dimensions. This indicates that no new divergencies appear, which we shall confirm next.

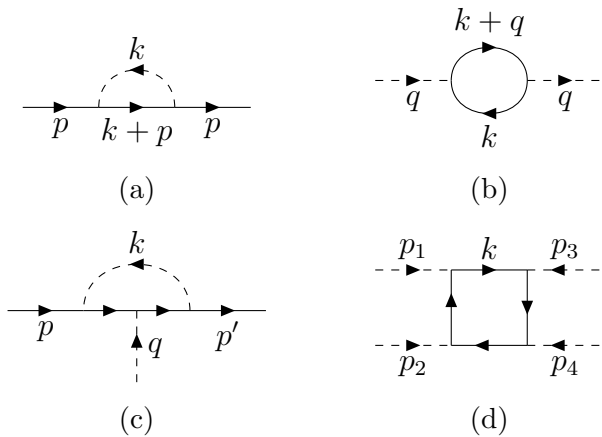
\begin{figure}[ht]
\centering
\begin{subfigure}{0.25\textwidth}
  \centering
  \feynmandiagram [small, layered layout, horizontal=b to c] {
    a -- [fermion, edge label'=\(p\)] b
      -- [fermion, edge label'=\(k+p\)] c
      -- [fermion, edge label'=\(p\)] d,
    c -- [charged scalar, half right, looseness=1.7, edge label'=\(k\)] b,
  };
  \caption{}\label{fig:2a}
\end{subfigure}
\hspace{1em}
\begin{subfigure}{0.25\textwidth}
  \centering
  \feynmandiagram [small, layered layout, horizontal=b to c] {
    a -- [charged scalar, edge label'=\(q\)] b
      -- [fermion, half left, looseness=1.5, edge label=\(k+q\)] c
      -- [charged scalar, edge label'=\(q\)] d,
    c -- [fermion, half left, looseness=1.5, edge label=\(k\)] b,
  };
  \caption{}\label{fig:2b}
\end{subfigure}
\\\vspace{.3em}
\begin{subfigure}{0.25\textwidth}
  \centering
  \begin{tikzpicture}
    \begin{feynman}[small]
      \vertex (a);
      \vertex [right=of a] (b);
      \vertex [right=7mm of b] (c);
      \vertex [right=7mm of c] (d);
      \vertex [right=of d] (e);
      \vertex [below=of c] (f);
      \diagram* {
        (a) -- [fermion, edge label'=\(p\)] (b)
        -- [fermion] (c)-- [fermion] (d)
        -- [fermion, edge label'=\(p'\)] (e),
      (c) -- [anti charged scalar, edge label=\(q\)] (f),
      (d) -- [charged scalar, half right, looseness=1.5, edge label'=\(k\)] (b),
      };
    \end{feynman}
  \end{tikzpicture}
  \caption{}\label{fig:2c}
\end{subfigure}
\hspace{1em}
\begin{subfigure}{0.25\textwidth}
  \centering
  \feynmandiagram [small, layered layout, horizontal=b to c] {
  a -- [charged scalar, edge label=\(p_1\)] b
  -- [fermion, edge label=\(k\)] c
  -- [anti charged scalar, edge label=\(p_3\)] d,
  e -- [charged scalar, edge label'=\(p_2\)] f
  -- [anti fermion] g
  -- [anti charged scalar, edge label'=\(p_4\)] h,
  { [same layer] b -- [anti fermion] f },
  { [same layer] c -- [fermion] g },
  };
  \caption{}\label{fig:2d}
\end{subfigure}

\caption{Diagrams of one-loop corrections for Yukawa interaction: (a) the fermion self-energy correction, (b) the scalar self-energy correction, (c) the fermion-scalar vertex correction, and (d) the vertex correction to interaction of four scalar fields; only one diagram is drawn, while the other box diagrams are obtained by permutations of the momenta $p_1,p_2,p_3,p_4$.
The diagrams with external scalar legs have different version for incoming and outgoing scalar particles due to Fig.~\ref{fig:treediagrams}.}
\label{fig:oneloopdiagrams}
\end{figure}

The UV divergent one-loop diagrams have the same structure as those of the local theory of Yukawa interaction. The one-loop diagrams are presented in Fig.~\ref{fig:oneloopdiagrams}. In the local theory, the integrals of all these one-loop diagrams have logarithmic divergencies. The question is how are the divergent loop integrals changed in the nonlocal theory \eqref{Lagrangian}.

First, consider the fermion self-energy correction, i.e., the correction to the fermion propagator,
\begin{equation}
S_F(p)\left[-i\Sigma(p)\right]S_F(p).
\end{equation}
The first contribution to the self-energy $\Sigma(p)$ is obtained from the one-loop diagram in Fig.~\ref{fig:1a} as
\begin{equation}\label{fermion-self-energy}
\begin{split}
-i\Sigma(p)&=\int\frac{d^4k}{(2\pi)^4}\,i^2\left\{ g^2+gg_1\left[ f_{+}(k)+f_{-}(k) \right] +g_1^2f_{+}(k)f_{-}(k) \right\} S_F(k+p)D_F(k)\\
&=\int\frac{d^4k}{(2\pi)^4}\left\{ g^2+gg_1\left[ f_{+}(k)+f_{-}(k) \right] + g_1^2f_{+}(k)f_{-}(k) \right\}\\
&\quad\times \frac{\slashed{k}+\slashed{p}+m-\mu \Delta f(k+p)}{(k+p)^2-[m-\mu \Delta f(k+p)]^2+i\epsilon}
\;\frac{1}{k^2-m_{\phi}^2+i\epsilon}.
\end{split}
\end{equation}

The self-energy correction for the scalar is obtained from Fig.~\ref{fig:2b} as
\begin{multline}\label{scalar-self-energy}
i\Pi(q)=-\int\frac{d^4k}{(2\pi)^4}\left\{ g^2+gg_1\left[ f_{+}(q)+f_{-}(q) \right] +g_1^2f_{+}(q)f_{-}(q)\right\}\\
\times\frac{\slashed{k}+m-\mu \Delta f(k)}{k^2-[m-\mu \Delta f(k)]^2+i\epsilon}
\; \frac{\slashed{k}+\slashed{q}+m-\mu \Delta f(k+q)}{(k+q)^2-[m-\mu \Delta f(k+q)]^2+i\epsilon}.
\end{multline}

The amputated 3-point function of the vertex correction is  obtained from Fig.~\ref{fig:2c} as
\begin{equation}\label{3-point-func}
\Gamma(p,p')=i\left[g+g_1f_+(p'-p)\right]\left[1-iV(p,p')\right],
\end{equation}
\begin{multline}\label{3-point-func.2}
V(p,p')=\int\frac{d^4k}{(2\pi)^4}\left\{ g^2+gg_1\left[ f_{+}(k)+f_{-}(k) \right] +g_1^2f_{+}(k)f_{-}(k)\right\} \\
\times\frac{\slashed{k}+\slashed{p}+m-\mu \Delta f(k+p)}{(k+p)^2-[m-\mu \Delta f(k+p)]^2+i\epsilon}\\
\times \frac{\slashed{k}+\slashed{p}'+m-\mu \Delta f(k+p')}{(k+p')^2-[m-\mu \Delta f(k+p')]^2+i\epsilon} 
\; \frac{1}{k^2-m_\phi^2+i\epsilon}.
\end{multline}

In all loop contributions, the difference compared to the local theory is the appearance of the sum $f_{+}(k)+f_{-}(k)$, the product $f_{+}(k)f_{-}(k)$ and the difference $\Delta f(k+\ldots)$ of the form factors \eqref{formfactors} in the nominator of the integrand of \eqref{fermion-self-energy} as well as the appearance of $\Delta f(k+\ldots)$ in the denominator of the integrand.

The form factors $f_{\pm}(k)$ go to zero in the limits $k^0\to\infty$ and $|\mathbf{k}|\to\infty$. The sum $f_+(k)+f_-(k)$, the difference $\Delta f(k)$ and the product $f_+(k)f_-(k)$ of the form factors share this property. Furthermore, in the integrands of the loop integrals, all contributions proportional to $g_1^b$ ($b=1,2,\ldots$) involve oscillating integrands due to the factors $e^{\pm ik\cdot z}$ in the form factors \eqref{formfactors}. As a result, in the limits  $k^0\to\infty$ and $|\mathbf{k}|\to\infty$, the contributions proportional to $g_1^b$ vanish due to the Riemann--Lebesgue lemma.
Thus, at high $k$ both the nominator and the denominator in \eqref{fermion-self-energy}--\eqref{3-point-func.2} behave similarly as in the local theory. Similar behavior is seen in all the other (higher order) loop integrals.
Therefore, the ultraviolet divergencies are of the same magnitude as in the corresponding local theory, and hence the nonlocal theory is renormalizable.

\section{Unitarity}

Unitarity in loop amplitudes can be considered with the Cutkosky--Landau cutting rules \cite{Landau:1959,Cutkosky:1960sp,Veltman:1963th}. Another method is the partial wave expansion, which has been widely used for obtaining bounds on the energy-dependence of cross sections \cite{Froissart:1961ux,Martin:1962rt} (generalized to processes with arbitrary number of particles \cite{Wang:1971cb} and higher dimensional spacetime \cite{Chaichian:1987zt,Chaichian:1992hq}), on the masses of Higgs boson  \cite{Dicus:1973gbw,Lee:1977yc,Lee:1977eg,Marciano:1989ns} and weakly interacting particles \cite{Chanowitz:1978uj,Chanowitz:1978mv}, and on parameters of many other models (e.g. on generic Yukawa couplings \cite{Allwicher:2021rtd}). Both of those methods are based on the quantum field theoretic formulation of the optical theorem.

For simplicity, we show that the unitarity is retained for unpolarized scattering. Let $\mathcal{M}(s,t)$ be the Lorentz invariant amplitude for scattering of two particles, where $s$ and $t$ are Mandelstam variables. The probability of unpolarized scattering is determined by the squared amplitude
\begin{equation}\label{unpolarized}
\frac{1}{(2s_1+1)(2s_2+1)}\sum_\mathrm{spins}\left|\mathcal{M}(s,t)\right|^2,
\end{equation}
where $s_1$ and $s_2$ are the spins of the initial particles and the sum is taken over the spin states of all initial and final particles. The first partial wave of \eqref{unpolarized}, which is defined as
\begin{equation}\label{A_0}
A_0(s)=\frac{1}{(2s_1+1)(2s_2+1)}\int_{-1}^1 d(\cos\vartheta)\sum_\mathrm{spins}\left|\mathcal{M}(s,t)\right|^2,
\end{equation}
where $\vartheta$ is the scattering angle, determines the total scattering probability, and consequently the total cross section, given in the center-of-mass frame as $\sigma=\frac{16\pi}{s}A_0(s)$.
Unitarity is ensured when \eqref{A_0} satisfies $A_0(s)\le\frac{1}{(2s_1+1)(2s_2+1)}$. For tree-level amplitudes, we impose a stricter condition:
\begin{equation}\label{A_0.tree-unitarity}
A_0^\mathrm{tree}(s)<\frac{1}{4(2s_1+1)(2s_2+1)}.
\end{equation}
Further details on the partial wave expansion and perturbative unitarity are presented in the Appendix, particularly on how the above condition ensures that the unitarity conditions for all partial waves \eqref{partialwave} are satisfied.

In the present theory, the unitarity condition \eqref{A_0.tree-unitarity} for every tree-level process has the following general form:
\begin{equation}\label{unitarity.general}
g^4F_0(s)+\sum_{b=1}^4 g^{4-b}g_1^bF_b(s) < \frac{(8\pi)^2}{(2s_1+1)(2s_2+1)},
\end{equation}
where in high energies the functions $F_b(s)$ reduce with energy as quickly as $s^{-b}$. Eq.~\eqref{unitarity.general} is a condition on the couplings $g$ and $g_1$, which is satisfied for sufficiently small values of the couplings.
In the high-energy limit, the contributions from the nonlocal interaction vanish, since a contribution proportional to $g_1^b$ scales with energy as $s^{-b}$. Thus, in the high-energy limit, the unitarity conditions reduce to those of the local theory, which have the form:
\begin{equation}
g^2 < \frac{8\pi}{\sqrt{(2s_1+1)(2s_2+1)}\sqrt{\lim_{s\to\infty}F_0(s)}},
\end{equation} 
where $\sqrt{\lim_{s\to\infty}F_0(s)}$ is a constant of the order of 1.

Furthermore, we see that higher-order (loop) contributions from the nonlocal interaction also scale as $g_1^b/s^b$, suggesting that the model remains unitary in all orders of perturbation theory.

Finally, we remark on the particle-antiparticle mass splitting ($\mu\neq0$). The mass term of the Dirac equation in momentum space is changed from $m$ to $m-\mu\Delta f(p)$. Since calculation of the partial wave expansion of scattering amplitudes without approximation appears very difficult in this case, we use the approximate canonical description \cite{Chaichian:2012bk}. We approximate the masses of particles and antiparticles by solving the eigenvalue equations in the rest frame,
\begin{equation}
p_0=\gamma_0\left( m\pm 4\pi\mu\int_{0}^{\infty}dz\frac{z^2\sin
[p_0\sqrt{z^2+l^2}]}{\sqrt{z^2+l^2}} \right),
\end{equation}
iteratively as
\begin{equation}\label{m_mp}
m_\pm\approx m\pm4\pi\mu\int_{0}^{\infty}dz\frac{z^2\sin
[m\sqrt{z^2+l^2}]}{\sqrt{z^2+l^2}},
\end{equation}
where $\mu$ is assumed to be sufficiently small so that the corrections to $m$ remain small, $|m_\pm-m|\ll m$. Then we treat fermions and antifermions as if they had the constant masses $m_+$ and $m_-$, respectively \cite{Chaichian:2012bk}. The analysis of unitarity then follows similarly as in the case with no mass splitting. The only major differences are that particles and antiparticles have different mass parameters and the fermion propagator is changed to \eqref{propagator.psi}. This does not jeopardize unitarity, as the dependence of the scattering amplitudes on $s$ and $t$ is not significantly altered. Nonlocal contributions to the amplitudes, which involve couplings with positive mass dimensions, are suppressed by powers of $g_1/s$ and $\mu/s^{3/2}$, compared to the local contributions. In the high-energy limit, mass splitting becomes inconsequential, since $\Delta f(p)\to0$.

In conclusion, the suggested nonlocal QFT satisfying all the requirements of a viable theory, can open a new path and be generalized to gauge theories as gauge invariance can be achieved in such a theory by the Schwinger phase factor \cite{Schwinger:1951nm}.
Most importantly, dressing such a nonlocal Standard Model with a Kobayashi--Maskawa CP violating phase, will make the theory also C- and CP-violating. The additional violation of CPT achieved here, does not only replace Sakhorov’s requirement \cite{Sakharov:1967dj} of being out of thermodynamical equilibrium, but also for the equilibrium case, since this violation prevents the detailed balance. Thus, most requirements for the baryon asymmetry of the universe can finally be explained by a perfectly viable QFT. As for the remaining requirement of baryon number violation, hopefully there are several ways to achieve it, typically in GUT and electroweak baryogenesis, leptogenesis or sphalerons. Further studies are certainly needed to accomplish the ideas presented in this Letter to reach the ultimate aim of explaining the baryon asymmetry in the universe.

\subsection*{Acknowledgments}
We are much grateful to Luis \'Alvarez-Gaum\'e, Michael D\"utsch and Claus Montonen for insightful discussions.

\appendix
\renewcommand{\theequation}{A.\arabic{equation}}
\setcounter{equation}{0}

\section{Appendix concerning perturbative unitarity}
Unitarity of the $S$-matrix, $S^\dagger S=1$, applied to the scattering amplitude $\mathcal{M}(i\to f)$, which is defined as $\braket{f|S-1|i}=i(2\pi)^4\delta^4(P_i-P_f)\mathcal{M}(i\to f)$, where $P_i$ and $P_f$ are the total four-momenta of the initial and final states, gives the generalized optical theorem:
\begin{equation}\label{opticaltheorem}
2\Imag\mathcal{M}(i\to f)=\sum_b\int d\Pi_b (2\pi)^4\delta^4(P_i-P_b)
\mathcal{M}^*(f\to b)\mathcal{M}(i\to b),
\end{equation}
where $b$ denotes all possible intermediate states, $d\Pi_b$ are the phase spaces of such states, and we have omitted the $\delta$-function which ensures energy-momentum conservation, $P_i=P_f$.

We expand the amplitude of scattering of two particles to two particles as \cite{Jacob:1959at}:
\begin{align}
\mathcal{M}(s,t,\lambda_n)&=16\pi\sum_J (2J+1)a_J(s,\lambda_n) d^J_{\bar\lambda_i\bar\lambda_f}(\vartheta),\label{amplitude}\\
a_J(s,\lambda_n)&=\frac{1}{32\pi}\int_{-1}^1 d(\cos\vartheta) \mathcal{M}(s,t,\lambda_n)d^J_{\bar\lambda_i\bar\lambda_f}(\vartheta).\label{partialwave}
\end{align}
The amplitude depends on $s=(p_1+p_2)^2=(p_3+p_4)^2$, $t=(p_1-p_3)^2=(p_4-p_2)^2$, where $p_n$ are the four-momenta of the initial ($n=1,2$) and final ($n=3,4$) particles, and on the helicities $\lambda_n$ of the initial and final particles. In the expansion \eqref{amplitude}, $a_J(s,\lambda_n)$ is the partial wave amplitude with total angular momentum $J$ and $d^J_{\bar\lambda_i\bar\lambda_f}(\vartheta)$ are Wigner $d$-functions with $\bar\lambda_i=\lambda_1-\lambda_2$ and $\bar\lambda_f=\lambda_3-\lambda_4$.
The $d$-functions are defined for the angular momentum states as \cite{Wigner}
\begin{equation}
d^j_{mm'}(\vartheta)=\bra{jm'}e^{-i\vartheta\hat J_y}\ket{jm}.
\end{equation}
Unitarity of the $S$-matrix \eqref{opticaltheorem} requires that each partial amplitude \eqref{partialwave} satisfies:
\begin{equation}\label{unitarity}
\left(\Real a_J\right)^2+\left(\Imag a_J-\frac{1}{2}\right)^2 \le\frac{1}{4}
\end{equation}
or
\begin{equation}
|a_J|\le1,\quad |\Real a_J|\le\frac{1}{2},\quad 0\le\Imag a_J\le 1. 
\end{equation}
For high-energy tree-level elastic amplitudes, which are real due to the optical theorem, the unitarity condition can be written
\begin{equation}\label{tree-unitarity}
|a_J^\mathrm{tree}(s,\lambda_n)|<\frac{1}{2},
\end{equation}
trusting that loop corrections produce a sufficient imaginary contribution to put the partial amplitude within the required circle in the complex plane \eqref{unitarity}.

To show unitarity, it is sufficient to consider unpolarized scattering, since we are not interested in which polarized scattering gives the largest partial amplitude but in ensuring that the conditions for unitarity \eqref{unitarity} are satisfied.
Therefore, we work with the unpolarized squared amplitude,
\begin{equation}
|\mathcal{M}|^2_\mathrm{unpol}=\frac{1}{(2s_1+1)(2s_2+1)}\sum_{\{\lambda_n\}}|\mathcal{M}|^2,
\end{equation}
where $s_1$ and $s_2$ are the spins of the initial particles and the sum is taken over the helicities of all initial and final particles, which determines the unpolarized scattering probabilities. We expand it as
\begin{align}\label{ampl2.unpol}
|\mathcal{M}|^2_\mathrm{unpol}&=(16\pi)^2\sum_\ell (2\ell+1)A_\ell(s)P_\ell(\cos\vartheta),\\
A_\ell(s)&=\frac{1}{2(16\pi)^2}\int_{-1}^1 d(\cos\vartheta)|\mathcal{M}|^2_\mathrm{unpol.} P_\ell(\cos\vartheta),\notag
\end{align}
where $P_\ell(\cos\vartheta)$ are Legendre polynomials. 
We can also expand $|\mathcal{M}|^2_\mathrm{unpol}$ in terms of the partial amplitudes $a_J$ using \eqref{amplitude} and
\begin{equation}
d^J_{\bar\lambda_i\bar\lambda_f}(\vartheta)d^{J'}_{\bar\lambda_i\bar\lambda_f}(\vartheta)=(-1)^{\bar\lambda_i-\bar\lambda_f} \sum_{\ell} \braket{J\bar\lambda_i J'-\bar\lambda_i|\ell 0}
\braket{J\bar\lambda_f J'-\bar\lambda_f|\ell 0}
P_\ell(\cos\vartheta),
\end{equation}
where $\braket{j_1m_1j_2m_2|jm}$ are Clebsch-Gordan coefficients. This gives $A_\ell$ in terms of $a_J$ as
\begin{multline}
A_\ell(s)=\frac{1}{(2s_1+1)(2s_2+1)} \sum_{J,J'} \frac{(2J+1)(2J'+1)}{(2\ell+1)}
\sum_{\{\lambda_n\}} a_J(s,\lambda_n)a_{J'}^*(s,\lambda_n) (-1)^{\bar\lambda_i-\bar\lambda_f}\\
\times\braket{J\bar\lambda_i J'-\bar\lambda_i|\ell 0}\braket{J\bar\lambda_f J'-\bar\lambda_f|\ell 0}.
\end{multline}
We only need the first term of the expansion \eqref{ampl2.unpol},
\begin{equation}\label{A_0.2}
A_0(s)=\frac{1}{(2s_1+1)(2s_2+1)}\sum_{J}(2J+1)\sum_{\{\lambda_n\}}|a_J(s,\lambda_n)|^2,
\end{equation}
which determines the total scattering probability.
If we require that
\begin{equation}\label{A_0.cond}
A_0(s)\le\frac{1}{(2s_1+1)(2s_2+1)},
\end{equation}
then every term in the sum over $J$ in \eqref{A_0.2} satisfies the condition
\begin{equation}
(2J+1)\sum_{\{\lambda_n\}}|a_J(s,\lambda_n)|^2\le 1.
\end{equation}
This ensures that $|a_J(s,\lambda_n)|\le1$ for all $J$ and $\lambda_n$, i.e., the unitarity condition for all partial waves is satisfied. When there are several nonvanishing partial amplitudes, the condition \eqref{A_0.cond} is stricter than necessary but always sufficient.
At the tree level we require that \eqref{A_0.tree-unitarity} is satisfied, so that \eqref{tree-unitarity} is satisfied.

As an example of the unitarity condition in the present theory \eqref{Lagrangian}, consider the scattering of a fermion and an antifermion. For simplicity, we turn off the mass splitting ($\mu=0$); the effect of mass splitting is discussed in the section concerning unitarity. The unpolarized squared amplitude is obtained as:
\begin{multline}\label{ampl.faf.unpol}
\frac{1}{4}\sum_\mathrm{spins}\left|\ %
\begin{tikzpicture}[baseline=(a)]
\begin{feynman}[small, inline=(a)]
\vertex (i1);
\vertex [below right=of i1] (a);
\vertex [right=of a, xshift=-8pt] (b);
\vertex [below left=of a] (i2);
\vertex [above right=of b] (f1);
\vertex [below right=of b] (f2);
\diagram* {
(i1) -- [fermion, edge label'=\(p_1\),near start] (a) -- [fermion, edge label'=\(p_2\),near end] (i2),
(a) -- [scalar] (b),
(f2) -- [fermion, edge label'=\(p_4\),near start] (b) -- [fermion, edge label'=\(p_3\),near end] (f1),
};
\end{feynman}
\end{tikzpicture}
\ +\ \begin{tikzpicture}[baseline=(c)]
\begin{feynman}[small, inline=(c)]
\vertex (i1);
\vertex [below right=of i1, yshift=10pt] (a);
\vertex [above right=of a, yshift=-10pt] (f1);
\vertex [below=of a, yshift=18pt] (c);
\vertex [below=of c, yshift=20pt] (b);
\vertex [below left=of b, yshift=10pt] (i2);
\vertex [below right=of b, yshift=10pt] (f2);
\diagram* {
(i1) -- [fermion, edge label'=\(p_1\)] (a) -- [fermion, edge label'=\(p_3\)] (f1),
(f2) -- [fermion, edge label'=\(p_4\)] (b) -- [fermion, edge label'=\(p_2\)] (i2),
a -- [scalar] b,
};
\end{feynman}
\end{tikzpicture}
\ \right|^2 \\
=\left|(g+g_1f_-(\sqrt{s}))(g+g_1f_+(\sqrt{s}))\right|^2\frac{(s-4m^2)^2}{(s-m_\phi^2)^2}\\
+\left|(g+g_1f_-(\sqrt{s}))(g+g_1f_+(\sqrt{s}))\right|(g+g_1f(t))^2
\frac{15m^4-4m^2(s+t)-st}{(s-m_\phi^2)(t-m_\phi^2)}\\
+(g+g_1f(t))^4\frac{(4m^2-t)^2}{(t-m_\phi^2)^2},
\end{multline}
where the form factors \eqref{formfactors} are given in center-of-mass frame as
\begin{align}
f_\pm(\sqrt{s})&=f_\pm(p_1+p_2)=\frac{2\pi}{s}\int_0^\infty dz\frac{z^2e^{\pm i\sqrt{z^2+sl^2}}}{\sqrt{z^2+sl^2}},\notag\\
f(t)&=f_\pm(p_1-p_3)=\frac{2\pi}{-t}\int_0^\infty dz\frac{z\sin z}{\sqrt{z^2-tl^2}}.\label{formfunction.t}
\end{align}
The first partial wave \eqref{A_0} of \eqref{ampl.faf.unpol} is obtained as:
\begin{multline}\label{A_0.faf}
(16\pi)^2 A_0^{\psi\Bar\psi\to\psi\Bar\psi}(s)
=\left|(g+g_1f_-(\sqrt{s}))(g+g_1f_+(\sqrt{s}))\right|^2
\frac{(s-4m^2)^2}{(s-m_\phi^2)^2} \\
+g^2\left|(g+g_1f_-(\sqrt{s}))(g+g_1f_+(\sqrt{s}))\right| \\
\times\left[ -\frac{s+4m^2}{s-m_\phi^2}
+\frac{(4m^2+m_\phi^2)s+4m^2m_\phi^2-15m^4}{(s-4m^2)(s-m_\phi^2)} \ln\left(1+\frac{s-4m^2}{m_\phi^2}\right) \right] \\
+\left|(g+g_1f_-(\sqrt{s}))(g+g_1f_+(\sqrt{s}))\right| \left[ 2gg_1G_1(s)+g_1^2G_2(s) \right] \\
+g^4 \left[1 +\frac{(m_\phi^2-4m^2)^2}{(s-4m^2+m_\phi^2)m_\phi^2} 
+\frac{2(4m^2-m_\phi^2)}{s-4m^2}\ln\left(1+\frac{s-4m^2}{m_\phi^2}\right) \right] \\
+4g^3g_1H_1(s)+6g^2g_1^2H_2(s)+4gg_1^3H_3(s)+g_1^4H_4(s),
\end{multline}
where we have introduced the following six integrals $G_b(s)$ $(b=1,2)$ and $H_b$ $(b=1,2,3,4)$:
\begin{equation}
G_b(s)=\lim_{\varepsilon\to0^+}\frac{1}{s-4m^2}\int_{-(s-4m^2)}^0 dt\,[f_\varepsilon(t)]^b\,\frac{15m^4-4m^2(s+t)-st}{(s-m_\phi^2)(t-m_\phi^2)},
\label{G_b}
\end{equation}
\begin{equation}
H_b(s)=\lim_{\varepsilon\to0^+}\frac{1}{s-4m^2}\int_{-(s-4m^2)}^0 dt\,[f_\varepsilon(t)]^b\,\frac{(4m^2-t)^2}{(t-m_\phi^2)^2},
\label{H_b}
\end{equation}
where $f_\varepsilon(t)$ is a regulated version of the form function \eqref{formfunction.t} 
\begin{equation}
f_\varepsilon(t)=\frac{2\pi}{(\varepsilon-t)}\int_0^\infty dz\frac{z\sin z}{\sqrt{z^2-tl^2}},
\end{equation}
which is required to perform the integrals over $t$ in \eqref{G_b} and \eqref{H_b}.
The results of those integrations are long expressions, which are not presented here. 
$G_b(s)$ and $H_b(s)$ have a standing similar to that of the form factors \eqref{formfactors} in the scattering amplitude, but instead for the first partial wave of the unpolarized squared amplitude. For high energies, $G_b(s)$ and $H_b(s)$ reduce with energy as $s^{-b}$.
The condition \eqref{A_0.tree-unitarity} that ensures tree-unitarity \eqref{tree-unitarity} is obtained from \eqref{A_0.faf} by imposing $A_0^{\psi\Bar\psi\to\psi\Bar\psi}(s)<\frac{1}{4^2}$, which has the form \eqref{unitarity.general}. In the high-energy limit, the condition reduces to $g^2<4\pi$.

Similar analysis is done for the other scattering processes, namely, the scattering of a fermion on a fermion, the annihilation of a fermion and an antifermion to two scalar particles, and the scattering of a scalar particle and a fermion. All the conditions obtained for unitary have the general form presented in \eqref{unitarity.general}.


\newcommand{\atitle}[1]{\emph{#1},}
\newcommand{\booktitle}[1]{\emph{#1}}
\newcommand{\ebook}[4]{\href{https://doi.org/#4}{\emph{#1}, #2, #3}}
\newcommand{\jref}[2]{\href{https://doi.org/#2}{#1}}
\newcommand{\arXiv}[2]{\href{http://arxiv.org/abs/#1}
{\texttt{arXiv:#1 [#2]}}}
\newcommand{\arXivOld}[1]{\href{http://arxiv.org/abs/#1}
{\texttt{arXiv:#1}}}

\end{document}